\documentclass[conference]{IEEEtran}
\usepackage{graphicx}

\ifCLASSINFOpdf
\else
\fi
\begin{document}
%
\title{Accelerating In-transit Isosurface Generation With Topology Preserving Compression}

\author{\IEEEauthorblockN{Yanliang Li}
\IEEEauthorblockA{
University of Alabama at Birmingham\\
Birmingham, AL 35294\\
yli3@uab.edu}
\and
\IEEEauthorblockN{Jieyang Chen}
\IEEEauthorblockA{University of Alabama at Birmingham\\
Birmingham, AL 35294\\
jchen3@uab.edu}
}

\maketitle

\begin{abstract}
Data visualization through isosurface generation is critical in various scientific fields, including computational fluid dynamics, medical imaging, and geophysics. However, the high cost of data sharing between simulation sources and visualization resources poses a significant challenge. This paper introduces a novel framework that leverages lossy compression to accelerate in-transit isosurface generation. Our approach involves a Compressed Hierarchical Representation (CHR) and topology-preserving compression to ensure the fidelity of the isosurface generation. Experimental evaluations demonstrate that our framework can achieve up to 4x speedup in visualization workflows, making it a promising solution for real-time scientific data analysis.
\end{abstract}


\IEEEpeerreviewmaketitle

\vspace*{-1em}

\section{Introduction}
Data visualization such as \textit{isosurface} generation is one of the most important routines that scientists use to extract insights from data. 
Isosurface generation is commonly used in several science domains such as computational fluid dynamics (CFD)~\cite{feltcher1988computational}, medical imaging~\cite{we1987marching}, geophysics~\cite{dupuy2008isosurface}, and meteorology~\cite{rautenhaus2012web}.
An isosurface represents a surface that consist of points of a constant \textit{isovalue} within a volume of space.
Given a domain significance value as isovalue, the position of an isosurface, as well as its relation to other neighboring isosurfaces, can provide clues to the underlying structure of a scalar field.
As visualizations are commonly used for monitoring scientific simulations or experiments for scientists to steering their scientific workflows in time, it is essential to build efficient visualization workflows that enables near real-time (NRT) feedback for users.

One major challenge of building efficient visualization workflows is reducing the cost of data sharing between data source (such as computing nodes for simulation or science instruments for experiments) and computing resource for visualization, since it would prohibitively expensive to share via file-based I/O~\cite{poeschel2021transitioning}. 
In-situ visualization that allow sharing data bypassing the filesystems to achieve much more responsive visualization has been proposed and developed for the past decade~\cite{ma2009situ}. 
For example, in-transit based in-situ~\cite{bennett2012combining} that streams data via network to remote computing facilities for visualization is a more viable option for workflows runs on interconnected science instruments.
However, with the advancement of scientific instruments and simulation codes, data are being generated with a dramatic increasing volume and velocity, which makes network-based data streaming the dominating bottleneck for in-transit visualization workflows.

Recently, many data reduction tools designed for scientific data have been proposed that can greatly reduce the data volume. 
Among those, error-controlled lossy compression is one the most effective reduction tools since they provide high compression ratios while controlling the lossy of accuracy of reconstructed data. 
When reducing large scale datasets, lossy compression often offers more effective data volume reduction compared with lossless compression since not every bytes of the original data are preserved.
To control the loss of information in data, several error controlled lossy compression methods have been proposed. Those compressors enable users to prescribe error bound at the compression time so that decompressed data respect such error bound relative to the original data. Depending on the reduction method used, these compressors can be classified as prediction-based (e.g., ISABELA~\cite{lakshminarasimhan2013isabela}, SZ~\cite{tao2017significantly, zhao2021optimizing}, and FPZIP~\cite{lindstrom2006fast}) and transform-based (e.g., ZFP~\cite{lindstrom2014fixed} and MGARD~\cite{ainsworth2018multilevel}).

Despite many data reduction techniques have been proposed, not much work has been done to use lossy compression to accelerate in-transit visualization. So, in this work we aim to develop an novel accelerated in-transit visualization framework existing leveraging lossy compression. As the initial stage of the project, we focus on adapting and optimizing lossy compression for isosurface generation.

\begin{figure}[t]
\centering
\vspace{-1em}
\includegraphics[width=1\columnwidth]{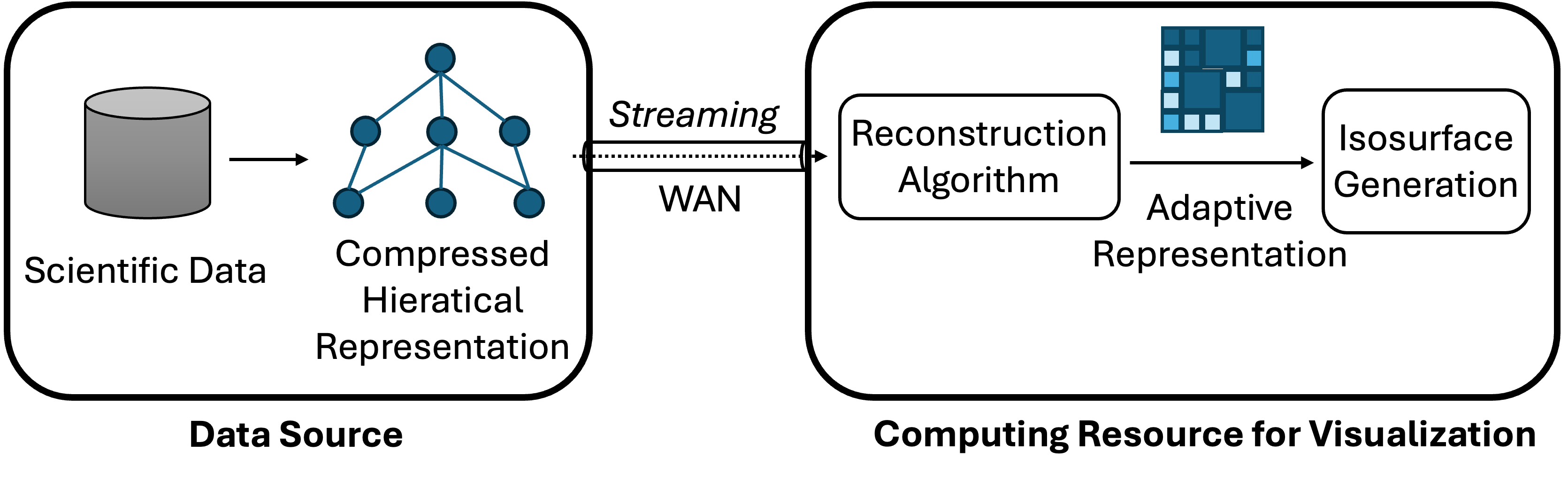}
\vspace{-2em}
\caption{Our accelerated in-transit isosurface generation framework}
\vspace{-2em}
\label{fig:framework}
\end{figure} 

\section{Compression Accelerated In-transit Isosurface Generation}
In this work, we focus on adapting and optimizing modern lossy compressors to in-transit isosurface generation workflows. 
Figure~\ref{fig:framework} shows our compression accelerated in-transit isosurface generation framework.
Left part represents the data source such as physical simulations running on large-scale high-performance computing (HPC) facilities or experimental instruments that generate data constantly.
The data will be pre-processed and compressed into a Compressed Hierarchical Representation (CHR).
On the visualization side (right), user can use our reconstruction algorithm to request data need to generate isosurface with desired isovalue and accuracy.

\subsection{Compressed Hierarchical Representation}
\begin{figure}[t]
\centering
\vspace{-1em}
\includegraphics[width=0.8\columnwidth]{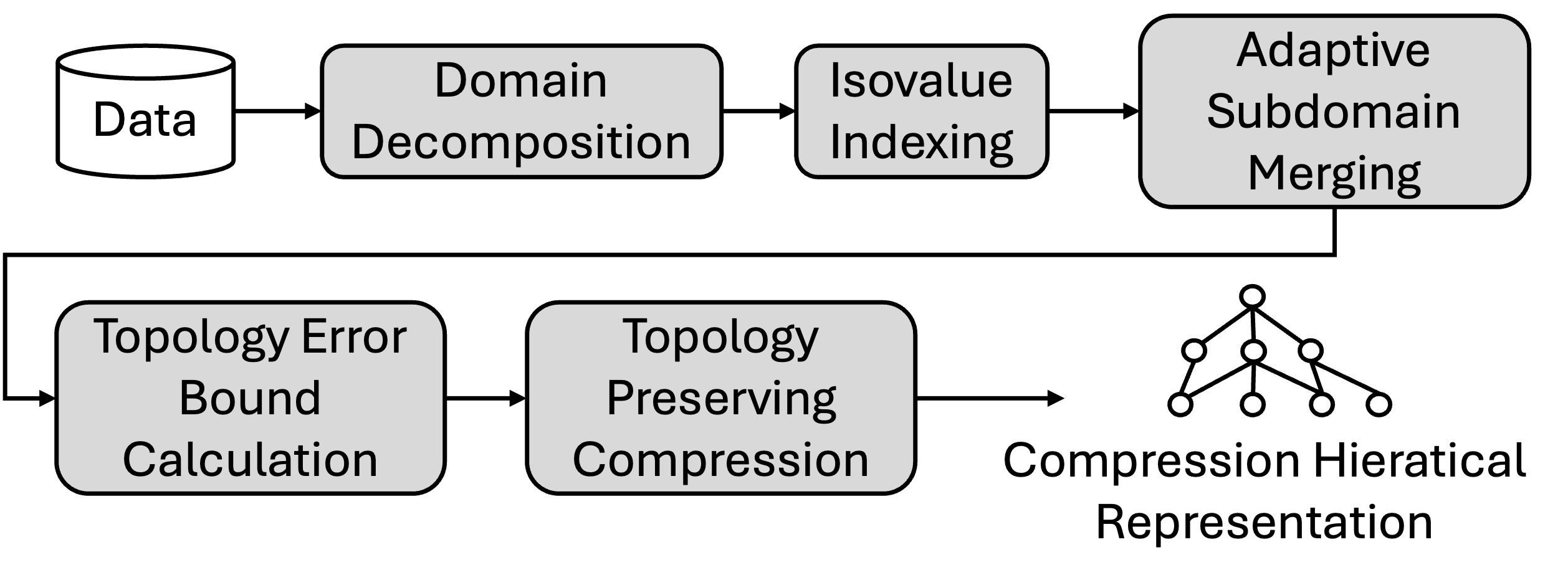}
\vspace{-1.5em}
\caption{Compressed Hierarchical Representation (CHR) generation process}
\vspace{-2em}
\label{fig:chr}
\end{figure}

Figure~\ref{fig:chr} illustrates the process of generating Compressed Hierarchical Representation (CHR).
We first apply domain decomposition by dividing data into blocks isovalue based data pruning and differentiated error control.
Then, to enable fast isosurface generation, we build index on blocks based on a series of candidate isovalues.
Next, we identify blocks that can be merge into larger rectangular grids so that they can be compressed and visualized more efficiently.
Finally, we apply our topology preserving compression to generate CHR.

\subsection{Topology Preserving Compression}

\begin{figure}[h]
\centering
\vspace{-2em}
\includegraphics[width=0.7\columnwidth]{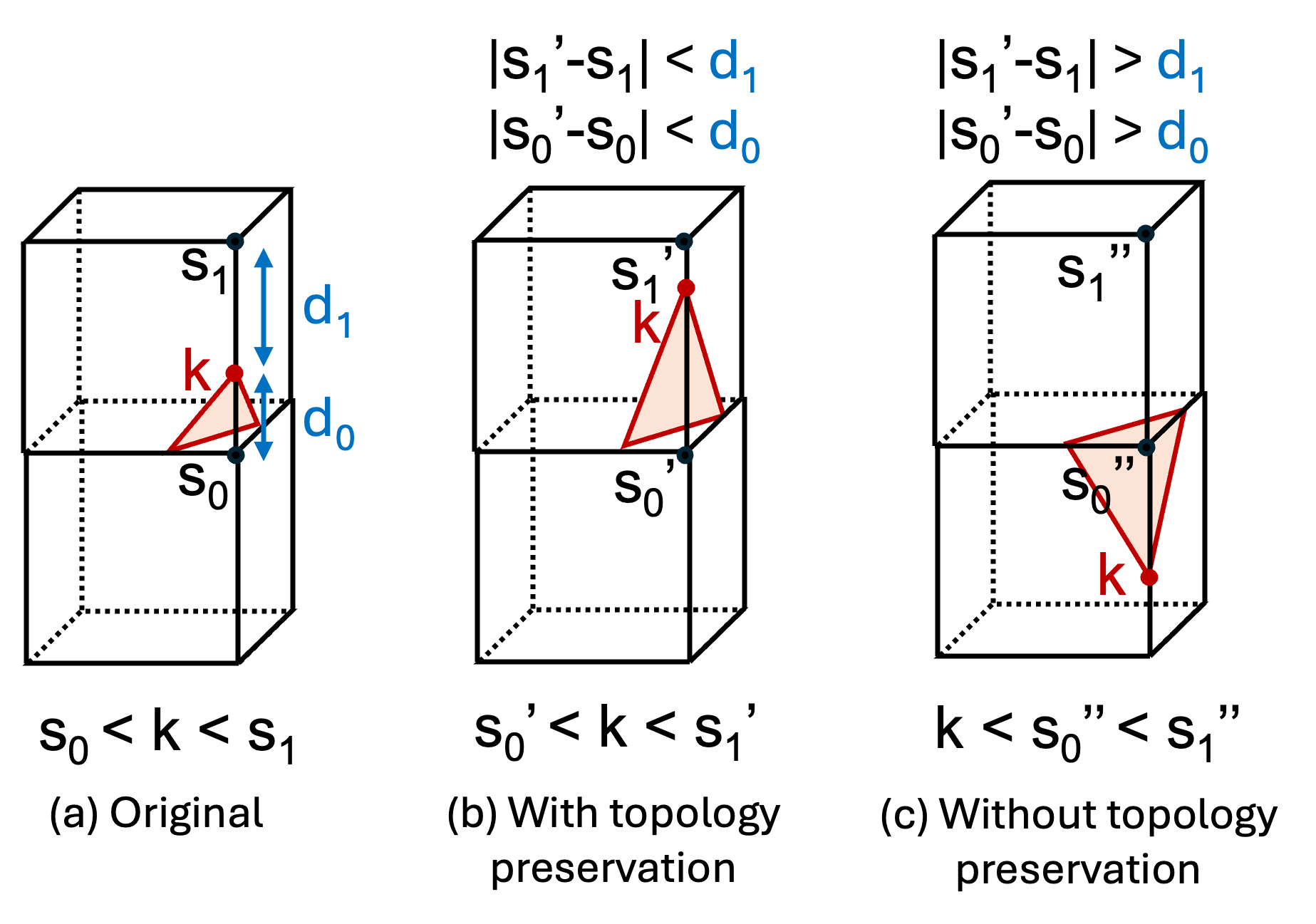}
\vspace{-1em}
\caption{Isosurface generation with and without topology preservation}
\vspace{-1em}
\label{fig:topology-preservation}
\end{figure} 

We propose to control the accuracy of isosurface generation by translating the error of isosurface to the error of raw data so that we can leverage existing lossy compressors to compress data while preserve the fidelity of isosurfaces.
To build such error translation, we first introduce the concept of \textit{topology preservation} for isosurfaces.
The isosurface generation process can be viewed as computing a surface out of a scalar field of the raw input data and the surface in each cell correspond to a certain case (i.e., cube configurations in term of Marching Cube~\cite{we1987marching}). So, we define topology preservation as guaranteeing that the same cases will be generated using decompressed data.
Figure~\ref{fig:topology-preservation} illustrates the process of generating isosurface in a cell with $k = isovalue$.
For the original data, $k$ falls in between the value of two vertexes $s_0$ and $s_1$.
If we incur error to $s_0$ and $s_1$, we might change the relationship between $s_0$, $s_1$, and $k$. 
On one hand, if the error is greater than the distance between $k$ and its two vertexes (i.e., $d_0$ and $d_1$), then case might change for the cell.
On the other hand, if the error is guaranteed to be smaller than both $d_0$ and $d_1$, we can guarantee that the case would remains the same.
So, to compute the error bound on raw data that preserve topology, we need to (1) find all edges ($s_0, s_1$) where $s_0 < k < s_1$; (2) compute $d_0$ and $d_1$ for all edges and save in an array $D$; (3) the smallest value in $D$ gives the absolute error bound for topology preserving compression. In addition, for cases where some topology errors are acceptable (e.g., tolerate erroneous 5\% cells), we can select the $n^{th}$ smallest in $D$ as error bound.

\section{Experimental Evaluation}
\begin{figure}[h!]
\centering
\vspace{-1em}
\includegraphics[width=1\columnwidth]{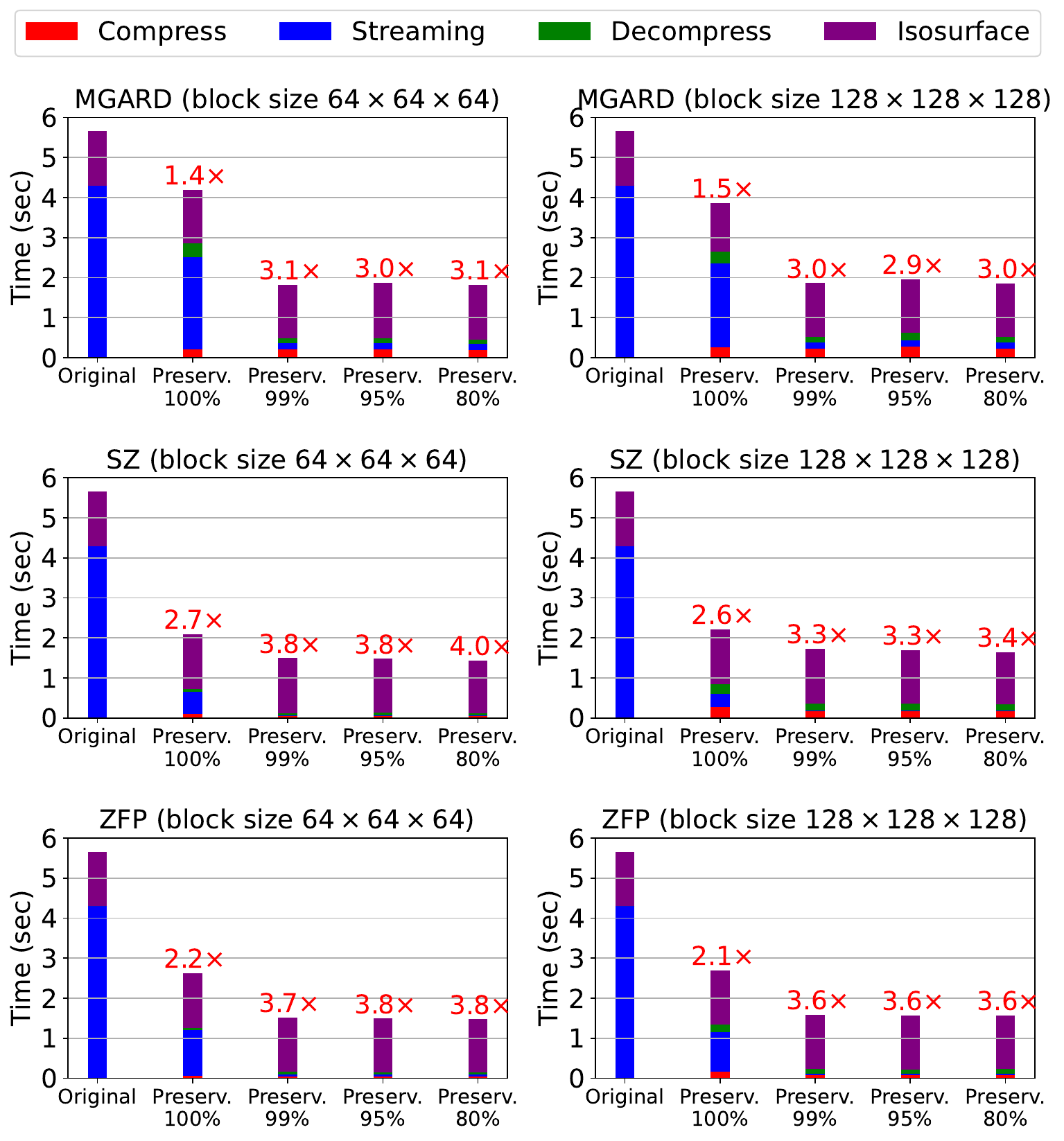}
\vspace{-2em}
\caption{End-to-end time breakdown of in-transit isosurface generation with and without our proposed topology preservation compression. Two block sizes ($64^3$ and $128^3$) for domain decomposition are tested.}
\vspace{-1em}
\label{fig:time-breakdown}
\end{figure}

We evaluate our framework on a compute node with Nvidia A100 GPUs on Jetstream 2 supercomputer~\cite{hancock2021jetstream2}~\cite{boerner2023access} at Indiana University. We use the temperature field data from NXY cosmological hydrodynamics simulation code~\cite{almgren2013nyx} for compression and isosurface generation. We simulate the data steaming process by calculating the streaming cost of transferring the original or compressed data over a 1Gpbs WAN. Figure~\ref{fig:time-breakdown} shows the end-to-end time breakdown of in-transit isosurface generation when using MGARD, SZ, and ZFP as compression backend. Comparing with the original workflow without compression, using 100\% preserving compression can achieve up to $2.6\times$ performance improvement. For cases that guarantee 99\%, 95\%, and 80\% topology accuracy, we can achieve up to $3.8\times$, $3.8\times$, $4.0\times$ speedups.

\section{Conclusion}
This work proposed a compression accelerated data streaming framework for in-transit visualization that are commonly used in scientific workflows with interconnected instruments. Although many lossy compression have been proposed with error control, none of them have been specifically designed for controlling the error of visualization output such as isosurface. We enabled topology error control by translating the error in topology to error into raw data. With our proposed framework, we can achieve up to $2.6\times$ speedup for topology error free isosurface generation and $4\times$ speedup for with error control.






\bibliographystyle{IEEEtran}
\bibliography{references.bib}



\end{document}